\documentclass[10pt]{amsart}
\usepackage{amssymb}
\usepackage[arrow,curve,all]{xy}
\def\intprod{\mathbin{\hbox to 6pt{%
                 \vrule height0.4pt width5pt depth0pt
                 \kern-.4pt
                 \vrule height6pt width0.4pt depth0pt\hss}}}

\let\hook\intprod
\def\intprod{\mathbin{\hbox to 6pt{%
                 \vrule height6pt width0.4pt depth0pt
                 \kern-.4pt
                 \vrule height0.4pt width5pt depth0pt\hss}}}

\let\lhook\intprod
\newcommand{\group}{$\mathfrak{G}$ }
\newcommand{\lalg}{$\mathfrak{g}$ }
\newcommand{\hft}{\Tilde{\mathcal{H}}^{n-1}(M,\Omega)}
\newcommand{\hftv}{\Tilde{\mathcal{H}}^*(M,\Omega)}

\newcommand{\hf}{\mathcal{H}^{n-1}(M,\Omega)}
\newcommand{\hfc}{\Check{\mathcal{H}}^{n-1}(M,\Omega)}

\newcommand{\isto}{\hspace{5pt} ::\hspace{5pt}} 
\newcommand{\hvf}{\text{Ham}(M,\Omega)}
\newtheorem{thm}{Theorem}[section]
\newtheorem{lem}[thm]{Lemma}

\newtheorem{prop}[thm]{Proposition}
\newtheorem{defn}[thm]{Definition}

\begin{document}

\title[On a Multisymplectic Formulation of the Classical BRST Symmetry...]
    {On a Multisymplectic Formulation of the Classical BRST Symmetry  for First Order Field Theories Part I: Algebraic Structures.
    }
\author{S.P.Hrabak.}
\address{Department of Mathematics\\
  King's College London\\ 
  Strand\\
   London WC2R 2LS\\
 England} 
\email{e-mail: shrabak@mth.kcl.ac.uk}
\thanks{Research supported by PPARC}

\subjclass{Primary: 53  Secondary: 14}
\date{January 20, 1999}


\begin{abstract}
We  describe a new algebraic multisymplectic formulation of the classical BRST symmetry. The analogue of Marsden-Weinstein reduction for multisymplectic manifolds is described. We then give  a homological description of Multisymplectic Marsden-Weinstein reduction.
\end{abstract}

\maketitle
The study of multisymplectic geometry \cite{Cans,CanG,CanH,Ibort,Martin}  arose in the context of the search for the geometric foundations of classical field theory \cite{Crampin,GIMMSY,Kij1,Kij2,KijSzc1,KijSzc2}. In this paper we shall formulate a homological description of Marsden-Weinstein multisymplectic reduction in the generic context of free and proper group actions on multisymplectic manifolds. In the paper which follows \cite{jdg2} we shall return to the specific context of those multisymplectic manifolds which form the geometric foundations of field theory.

Underlying the well known Hamiltonian (n+1)\footnote{n+1 is the dimension of the ambient space-time. The (n+1) denotes a decomposition of space-time into space and time, which breaks manifest covariance.} formulation of classical field theory  is the more fundamental and  covariant framework known as the  multisymplectic formalism. In the multisymplectic formalism one considers a finite number of fields at a given event of  space-time in distinction to the (n+1) formalism where one considers an  infinity of fields at a given instant of time. Recently much progress has been made in elucidating the  ground and source of many of the  commonalities of constrained dynamical systems as observed within  the (n+1) formalism by studying the intrinsic structures of such theories in the multisymplectic framework. In particular, sufficient conditions for much of presumed characteristics of classical field theories within the Dirac-Bergman formalism have been shown to arise out of a  general study of classical field theory and especially covariant momentum maps within the multisymplectic formalism \cite{GIMMSY}. Also, recent applications have shown that multisymplectic structures provide a natural setting for the study of dispersive wave propagation problems and is useful in the study of periodic pattern formation and wave instability \cite{MarShk}. 

In the study of classical mechanical physical systems  the trajectories of particles are embeddings of the real line into a target space. The space of all equivalent curves at a given point in the target space then leads to the notion of the tangent bundle and its dual the cotangent bundle. The trajectories of particles are thus one-parameter curves where we identify the parameter as  time. In recent studies of fundamental  physics the notion of ``particle'' $\Longleftrightarrow$ ``curve'' as fundamental dynamical objects has been extended  by the notion of ``p-brane'' $\Longleftrightarrow$ ``submanifold''. Corresponding to the single parameter of curves are several parameters, the intrinsic (local) coordinates on the embedded manifold. There is therefore no obvious choice of single evolutionary parameter, that is time. By making a choice of a hypersurface one is however able to enforce a distinction and treat the system as an infinite dimensional classical mechanical system. Then so long as any constraints of such a theory are describable in terms of functions of the derivatives of the induced coordinates with respect to the single time parameter of the (n+1) formalism, the generalised Dirac-Bergman Hamiltonian formalism may be applied to their study. However one can imagine a more generic situation  where a defining constraint on a class of embeddings is such that the constraints  depend quintessentialy on the derivatives of the induced coordinates with respect to all of the several independent variables. An example of such a constraint is that defining the embedding of a pseudoholomorphic curve into an almost complex manifold. It is therefore timely that much recent effort has been in progress in developing a framework in which field theories  may be covariantly studied. With these considerations in mind it seems reasonable to direct ones attention towards formulating some of the more powerful techniques of the (n+1) formalism within the covariant framework of multisymplectic geometry. 

In this and an accompanying paper \cite{jdg2} we shall develop a covariant Hamiltonian formulation of the classical BRST symmetry. In this first paper we shall detail  a homological description of the multisymplectic analogue of Marsden-Weinstein reduction, generalising the known results from symplectic to multisymplectic geometry. A fortiori, we thus generalise the known results from the symplectic context of classical mechanics to the multisymplectic formulation of first order field theory. This is necessary in order that the geometric framework of the second paper \cite{jdg2}  should have meaning. An outline of this paper is as follows:
\begin{enumerate}
\item In the first section we shall briefly review  the geometric and algebraic structures  of  multisymplectic manifolds prerequisite to the exposition in this paper. This review  blends together the approaches taken within  a number of seminal papers \cite{Cans,CanG,CanH,Crampin,GIMMSY,kan1,kan2,kan3,MarShk}. In particular we use the  graded Poisson-Leibniz algebra of observables of \cite{kan1,kan2,kan3}.
\item In the second section we describe the analogue of Marsden-Weinstein symplectic reduction for multisymplectic manifolds. We call this Marsden-Weinstein multisymplectic reduction. The  treatment given here was made possible by the treatment of multisymplectic reduction in \cite{CanG}.
\item In the third section we describe the dual process of obtaining the reduced observables of the reduced multisymplectic submanifold by the methods of homological algebra. We follow closely the general structure of the formulation for symplectic manifolds as given in \cite{KS}. The  added complexity of the algebra of observables being a graded Poisson-Leibniz algebra leads to some interesting and rather novel structures.
\item In the final section we conclude with a few remarks.
\end{enumerate}

\section{A Brief Review of the Multisymplectic Formalism}

All of the material in the section is taken from the literature, namely  \cite{Cans,CanG,CanH,Crampin,GIMMSY,kan1,kan2,kan3,MarShk} to which the reader is recommended for further details and proofs.

A {\it{multisymplectic manifold}} is a pair $(M,\Omega)$ where $M$ is a (m+n)-dimensional differentiable manifold and $\Omega \varepsilon \Omega^{(n+1)}(M)$ is 1-nondegenerate and closed. Such an $\Omega$ is called a {\it multisymplectic form}. The {\it 1-nondegeneracy} condition means that the  pairing $\Omega : T_xM \longrightarrow \Lambda^{n}(T^*_xM)
\hspace{10pt}\forall x \varepsilon M$ defines a bundle monomorphism.
Equivalently $\Omega$ is {\it 1-nondegenerate} iff $X \hook \Omega=0 \Leftrightarrow X=0$ for $X$ a vector field on $M$, or finaly $\Omega$ is 1-nondegenerate iff it has zero characteristic distribution. When $\Omega$ is globally exact so that $\Omega=-d\Theta$ then $\Theta$ is called the {\it multisymplectic potential}.When  n+1=2 we regain the definition of a symplectic manifold. For n+1=3 m+n=6 a Calabi-Yau 3-fold is an example of a multisymplectic manifold, for it is endowed with a 1-nondegenerate holomorphic 3 form. For first order field theories the affine dual of the first jet bundle over the bundle whose sections are the fields of interest is a multisymplectic manifold \cite{MarShk,Crampin,GIMMSY}. 

 A diffeomorphism $\Phi:M \longrightarrow M$ is said to be a {\it{multisymplectomorphism}} iff it preserves the multisymplectic form. That is, iff
$\Phi^*\Omega=\Omega$. It is called a {\it special multisymplectomorphism} iff it preserves the multisymplectic potential (for those multisymplectic manifolds for which such a potential exits globally). A vector field $X$ is said to  be {\it locally Hamiltonian} iff $ \mathcal{L}_X \Omega =0$, ie. iff it gives rise to flows which are (local) multisymplectomorphisms. A vector field $X$ is  {\it globally Hamiltonian}, or simply {\it Hamiltonian} iff there exits a (n-1)-form $F$ such that $X\hook\Omega=dF$ and $F$ is called a {\it Hamiltonian (n-1)-form}. The set of Hamiltonian vector fields shall be denoted $ \text{Ham}(M,\Omega)$ and set of Hamiltonian (n-1)-forms shall be denoted $\hf$. That is,
\begin{equation*}\hf:=\{F\varepsilon \Omega^{n-1}(M) : X\hook \Omega = dF \text{ for some vector field } X\}
\end{equation*}
Because $\Omega$ is 1-nondegerate the set of characteristic vector fields is zero. Given a Hamiltonian form F there is therefore a unique vector field corresponding to it, and we denote it by $X_F$. We label the map defining the correspondence by X. The correspondence between Hamiltonian (n-1)-forms and vector fields is thus well defined. Two Hamiltonian forms which differ by a closed n-1 form define the same vector field. The correspondence between vector fields  and Hamiltonian n-1 forms is thus well defined modulo closed (n-1)-forms. Let $\hft$ denote the {\it algebra of Hamiltonian forms on $M$ modulo closed  n-1 forms}. It is therefore clear that $\hft \cong \hvf$.  Let $\hfc$ denote the set of Hamiltonian forms modulo exact forms. 

The {\it{ bracket}} of two Hamiltonian forms is defined by $ \{F,G\}:=-X_F\hook X_G \hook \Omega $\footnote{N.B. the space $\hft$ is closed under the bracket as may be verified from the identity \(-X_F\hook X_G \hook \Omega=d(-X_F \hook X_G\hook\Omega)=:d\{F,G\}\), which follows by virtue of the Cartan's formula. Similar results hold for the other algebras introduced below. See refs. \cite{kan1,kan2,kan3}.}. Let $F,G,H$ be  Hamiltonian (n-1)-forms. They satisfy the following identities: (i) $[X_F,X_G] \hook \Omega =X_{\{F,G\}}\hook \Omega $, and (ii) $\{\{F,G\},H\} +cyclic=d(X_F\hook X_G\hook X_H \hook \Omega)$. It then follows that  $(\hfc,\{\cdot,\cdot\})$ and  $(\hft,\{\cdot,\cdot\})$ are  Lie Algebras. 

An (n-p)-vector field $\overset{n-p}{X}\varepsilon\Gamma(M,\Lambda^{n-p}(TM))$ is  {\it globally Hamiltonian}, or simply {\it Hamiltonian} iff there exits a p-form $\overset{p}{F}$ such that $\overset{n-p}{X}\hook\Omega=d\overset{p}{F}$ for $0\leq \text{p} \leq \text{n-1}$. Such a  $\overset{p}{F}$ is called a {\it Hamiltonian p form}. The set of Hamiltonian multi-vector fields modulo characteristic multi-vector fields shall be denoted $ \text{Ham}^*(M,\Omega)$ and set of Hamiltonian  forms modulo closed forms shall be denoted $\hftv$. That is, $\hftv:=\{F\varepsilon \Omega^*(M) : X\hook \Omega = dF $ for some multi-vector field $X\varepsilon\text{Ham}^*(M,\Omega)\}$. $\text{Ham}^*(M,\Omega)$ and $\hftv$ are isomorphic as vector spaces, where unlike the case of Hamiltonian vector fields there exist non-trivial characteristic multivector fields which we must  quotient out in order to obtain the bijective correspondence. If M is connected then $\Tilde{\mathcal{H}}^0(M,\Omega)\cong \frac{C^\infty(M,\Omega)}{\mathbb{R}}$. The {\it{ bracket}} of two Hamiltonian forms is defined by
\begin{equation} \{\overset{p}{F},\overset{q}{G}\}:=(-1)^{n-p}\overset{n-p}{X}_F\hook \overset{n-q}{X}_G \hook \Omega=:(-1)^{n-p}\overset{n-p}{X}_F\hook d \overset{q}{G}
\end{equation}
From this definition we learn that $(\hftv,\{\hspace{3pt},\hspace{3pt}\})$ is a graded Lie algebra with:
\begin{enumerate}
\item Graded bracket
\begin{equation}\{\overset{p}{F},\overset{q}{G}\}=-(-1)^{g_F g_G}\{\overset{q}{G},\overset{p}{F}\}\end{equation}
\item Graded Jacobi Identity
\begin{equation}(-1)^{g_F g_H}\{\{F,G\},H\} +(-1)^{g_F g_G}\{\{G,H\},F\} +(-1)^{g_G g_H}\{\{H,F\},G\}=0\end{equation}
\end{enumerate}
where $g_F=n-1-|F|$,$g_G=n-1-|G|$,$g_H=n-1-|H|$ and $|F|,|G|,|H|$ denote the form degrees of F,G and H.

Although it is common in the literature to refer to the above graded Lie bracket on the space of Hamiltonian forms as a Poisson bracket this is manifestly not the case as we have not introduced a  Poisson structure at this point. We recall that a {\it Poisson algebra} is an  associative algebra $\mathcal{A}$ over a field $\mathbb{K}$ carrying a Lie algebra bracket $\{\cdot,\cdot\}$ for which each adjoint operator $X_F=\{\quad,F\}$ is a derivation on the associative algebra structure. In \cite{kan2} I.V.Kanatchikov proposes a graded Leibniz algebra with a graded derivation  as a generalisation for field theory of the Poisson algebra of observables in classical mechanics (see also \cite{kan1,kan3}). A Leibniz algebra is a non-commutative generalisation of Lie algebra introduced by J-L.Loday\cite{loday}.  A {\it Leibniz algebra} over the field $\mathbb{K}$ is a vector space $\mathcal{L}$ equipped with a binary operation $[-,-]:\mathcal{L}\otimes\mathcal{L}\rightarrow \mathcal{L}$ satisfying the {\it Loday identity}
$[[x,y],z]=[x[,y,z]]-[y,[x,z]]$.  We call such a binary operation a {\it Leibniz bracket}. We shall call an associative algebra over a field $\mathbb{K}$ carrying a Leibniz bracket for which each adjoint operator is a (right or left) derivation on the associative algebra structure a {\it Poisson-Leibniz algebra}. The  special case of a Poisson-Leibniz algebra for which the bracket is (graded) anti-commutative is a (graded) Poisson(-Lie)  algebra.

We need to build from the graded Lie algebra $\hftv$ an associative algebra structure. The most natural candidate is to consider the exterior product on forms. However space of Hamiltonian forms is not stable under the exterior product. That is, the exterior product of two Hamiltonian forms  $\overset{p}{F}\wedge\overset{q}{G}$ is not generically Hamiltonian. It is possible however by extending the notion of Hamiltonian multi-vector fields to form-valued multi-vector fields to make sense of $\overset{p}{F}\wedge\overset{q}{G}$ as a ``generalised'' Hamiltonian form in $\Lambda^*(\hftv)$\footnote{N.B. the exterior algebra over the graded vector space $\hftv$ is defined by quotienting out the tensor algebra over $\hftv$ by the ideal generated by the relations $\mathfrak{F}^p\otimes\mathfrak{H}^q -(-)^{pq}\mathfrak{H}^q\otimes\mathfrak{F}^p$ for all Hamiltonian forms $\mathfrak{F}^p$, $\mathfrak{H}^q$.} \cite{kan1,kan2,kan3}. Given the exterior product of two Hamiltonian forms  $\overset{p}{F}\wedge\overset{q}{G}$ the corresponding form-valued multi-vector field is determined by a generalised multisymplectic structural equation, $\Check{X}_{\overset{p}{F}\wedge\overset{q}{G}}\hook\Omega=:d(\overset{p}{F}\wedge\overset{q}{G})$,  to be 
\begin{equation}\Check{X}_{\overset{p}{F}\wedge\overset{q}{G}}=(-1)^{(p+1)q} \overset{q}{G}\wedge X_{\overset{p}{F}} + (-1)^p\overset{p}{F}\wedge X_{\overset{q}{G}}.\end{equation}
Note that the generalised multisymplectic structural equation implies that \[\mathcal{L}_{\Check{X}_{\overset{p}{F}\wedge\overset{q}{G}}}\Omega=0.\]
The {\it{ Leibniz bracket}} of two generalised Hamiltonian forms
$\overset{r}{\mathfrak{F}},\overset{s}{\mathfrak{G}}\varepsilon\Lambda^*(\hftv)$ is defined by
\begin{equation} \{\overset{r}{\mathfrak{F}},\overset{s}{\mathfrak{G}} \}:=(-1)^{n-r}\Check{X}_{\mathfrak{F}}\hook d\overset{s}{\mathfrak{G}}. \end{equation}
The bracket on $\Lambda^*(\hftv)$ so defined is non-commutative as may be verified by computing and contrasting $\{\mathfrak{F}\wedge \mathfrak{G},\mathfrak{H}\}$ and $\{\mathfrak{H},\mathfrak{F}\wedge \mathfrak{G}\}$. The algebraic structure defined by $(\Lambda^*(\hftv),\{\cdot,\cdot\})$ does not possess a (graded nor ungraded) Lie algebra structure. 

The {\it algebra of observables} is the graded Poisson-Leibniz algebra \[(\Lambda^*(\hftv),\{\cdot,\cdot\})\] satisfying the following  identities:
\begin{enumerate}
\item The {\it left graded Loday identity}
\begin{equation}\{\{\mathfrak{F},\mathfrak{G}\},\mathfrak{H}\}=\{\mathfrak{F},\{\mathfrak{G},\mathfrak{H}\}\}-(-1)^{g_\mathfrak{F} g_\mathfrak{G}}\{\mathfrak{G},\{\mathfrak{F},\mathfrak{H}\}\}\end{equation}
\item The {\it right graded Leibniz rule}
\begin{equation}\{\mathfrak{F}\wedge \mathfrak{G},\mathfrak{H}\}=\mathfrak{F}\wedge\{\mathfrak{G},\mathfrak{H}\} +(-1)^{|\mathfrak{G}|(g_\mathfrak{H})}\{\mathfrak{F},\mathfrak{H}\}\wedge \mathfrak{G}\end{equation}
\end{enumerate}

We shall advocate this graded Poisson-Leibniz algebra of observables as the field theoretic generalisation of the Poisson algebra of observables in classical mechanics. Its introduction enables us to give a homological description of Marsden-Weinstein multisymplectic reduction and is essential to the geometrisation of the algebraic structures of this paper in \cite{jdg2}.

Let \group be a Lie group with Lie algebra \lalg, and let $ \Phi:$\group$ \times M \longrightarrow M$ be a multisymplectic action of \group  on $M$.
Let $\eta_M$ denote the vector field on $M$ associated with the Lie algebra element $\eta$ by virtue of this group action. We shall assume that the action is such that the vector field $\eta_M$ is a Global Hamiltonian vector field so that there exists a corresponding Hamiltonian form $F_{\eta}$. The map $\phi$, see the diagram below, sending $\eta$ to $\eta_M$ is a Lie algebra homomorphism. However the map $\delta$ from \lalg to $\hfc$ is not generically a Lie algebra homomorphism for  $\{F_{\eta},F_{\zeta}\}=F_{[\eta,\zeta]} + c(\eta,\zeta)$ where $c(\eta,\zeta)$ is a 2-cocycle on \lalg with values in $H^{n-1}(M)$ \cite{Crampin}. When this  2-cocycle does not vanish one regains the homomorphism property by forming a central extension to the group.
\[
\xymatrix{
&&& {\mathfrak{g}}  \ar[d]^{\phi} \ar[dl]_\delta \\
0 \ar[r] & H^{n-1}(M) \ar[r]^i &\hfc  \ar[r]^X& Ham(M,\Omega)
\ar[r]& 0
}
\]
The {\it covariant multimomentum map} $\mathbb{J}$ for the action of \group on $M$ is the map $\mathbb{J}:M \longrightarrow $\lalg$^*\otimes \Lambda^{n-1}M$
such that $d\delta(\eta)=\eta_M \hook \Omega $ for all $\eta \varepsilon $\lalg, where $\delta(\eta)(m)=:<\eta,\mathbb{J}(m)>$. We call $\delta(\eta)$ the {\it covariant Noether current} corresponding to the Lie algebra element $\eta$. The covariant multimomentum map satisfies the following equivariance condition: for any,  $g \varepsilon$\group $g^*<\eta,\mathbb{J}>=<\text{Ad}g^{-1}\eta,\mathbb{J}>$ (modulo closed forms). This  is equivalent to the commutivity of the following diagram
\[
\xymatrix{
M  \ar[d]_g \ar[r]^{\mathbb{J}} & {\mathfrak{g}}^*\otimes \Lambda^{n-1} M  \ar[d]^{{\text{Ad}_g}^*\otimes {\text{Id}}} \\
M \ar[r]^{\mathbb{J}} & {\mathfrak{g}}^* \otimes \Lambda^{n-1} M 
}
\]

\section{Marsden-Weinstein Multisymplectic Reduction}
In this section we describe the analogue of Marsden-Weinstein reduction for symplectic manifolds in the multisymplectic case. Marsden-Weinstein reduction is a special case of symplectic reduction corresponding to  a group acting by symplectomorphisms on a symplectic manifold. Symplectic reduction is based on the observation that every coisotropic submanifold of a symplectic manifold is foliated by isotropic leaves and that the quotient, if it exists, is again a symplectic manifold \cite{dusa}. The analogue of this observation for multisymplectic manifolds was obtained in \cite{CanG}. In Marsden-Weinstein reduction the coisotropic submanifold is the zero set of the moment map, and the isotropic leaves are the orbits of the group. Let \group be a Lie group with Lie algebra \lalg, and let $ \Phi:$\group$ \times M \longrightarrow M$ be a multisymplectic action of \group  on $M$. Let the covariant momentum map be $\mathbb{J}$. Recall that  the moment map is \group-equivariant. We may therefore conclude, that since 0$\varepsilon$\lalg is a fixed point of the coadjoint action, that $\mathbb{J}^{-1}(0)$ is invariant under the action of $\mathfrak{G}$. We assume the following:
\begin{enumerate}
\item  0 is a regular value of $\mathbb{J}$ so that $\mathbb{J}^{-1}(0)$ is a submanifold of $M$. 
\item The group  $\mathfrak{G}$ acts freely and properly on $\mathbb{J}^{-1}(0)$ so that the quotient $\frac{\mathbb{J}^{-1}(0)}{\mathfrak{G}}$ is a manifold.
\end{enumerate}
 The submanifold defined by the vanishing of the covariant Noether currents $\mathbb{J}^{-1}(0)=:\mathcal{C}$ is  the {\it constraint submanifold} and the quotient 
$\frac{\mathbb{J}^{-1}(0)}{\mathfrak{G}}=:\mathcal{B}$ the {\it reduced multisymplectic manifold}. Let $\mathfrak{j}$ be the surjective mapping  of $\mathcal{C}$ onto $\mathcal{B}$.
Under these assumptions and given the results of  \cite{CanG} we have the following analogue of the Marsden-Weinstein Reduction Theorem.

{\begin{thm} $\mathcal{C}$ is a coisotropic submanifold of the multisymplectic manifold $(M,\Omega)$, and the  leaves of the corresponding isotropic foliation are given by the orbits of $\mathfrak{G}$. The quotient $\mathcal{B}$ is again  multisymplectic manifold with a unique multisymplectic form $\Acute{\Omega}$ such that 
$\mathfrak{j}^*({\Acute{\Omega}})=\mathfrak{i}^*(\Omega)$ where $\mathfrak{i}$ denotes the embedding of $\mathbb{J}^{-1}(0)$ in $M$.
\end{thm}}

See reference \cite{CanG} for a discussion of isotropic, co-isotropic, lagrangian and multisymplectic submanifolds of multisymplectic manifolds.
\section{Homological Description of Marsden-Weinstein Multisymplectic Reduction}
In the following two sections we will  formulate a homological treatment of Marsden-Weinstein multisymplectic reduction analogous to the treatment of Marsden-Weinstein symplectic reduction given in \cite{KS}. The homological description of the reduction process takes place in two parts. In the first subsection  we obtain the observables on $\mathcal{C}$. In the second we obtain the reduced observables on $\mathcal{B}$. From our knowledge of symplectic reduction we would expect the observables on $\mathcal{C}$ to be the observables on $M$ modulo the ideal generated by the covariant Noether currents $\delta[\mathfrak{g}]$. That is,
\begin{equation}
\Lambda^*(\mathfrak{\Tilde{\mathcal{H}}^*(\mathcal{C})})
\cong \frac{\Lambda^*(\hftv)}
{\Lambda^*(\hftv)\wedge \delta[\mathfrak{g}]}
\end{equation}
This result is the content of Theorem 3.6.
Furthermore the observables on $\mathcal{B}$ ought to be those observables on $\mathcal{C}$ which are $\mathfrak{g}$-invariant. This result is the content of Theorem 3.12.

\subsection{The Koszul Complex}
In this subsection we introduce a certain distribution over $M$ where each fibre of the distribution carries a representation of the Lie algebra $\mathfrak{g}$. This representation possesses a  grading property which will allow  us to construct a Koszul complex from the basis spanning each fibre. The resulting homology   yields a resolution of the algebra of observables on the constraint submanifold. 

Let $\{\xi_a\}_{a=1 \cdots dim \mathfrak{g}}$ be a basis of the Lie algebra $\mathfrak{g}$ and $\{ X(\xi_a) \}_{a=1 \cdots dim \mathfrak{g}}$ the corresponding basis of Hamiltonian vector fields and $\{ \delta(\xi_a) \}_{a=1 \cdots dim \mathfrak{g}}$ the Hamiltonian (n-1)-forms determined from $X(\xi_a)\hook\Omega=d\delta(\xi_a)$. The $\{ X(\xi_a) \}_{a=1 \cdots dim \mathfrak{g}}$ (i) span the tangent space to the coisotropic submanifolds (diffeomorphic to $\mathfrak{G}$) foliating $\mathcal{C}$ and (ii) carry a representation of $(\mathfrak{g},[-,-])$ since $[X(\xi_a),X(\xi_b)]=C^d_{ab}X(\xi_d)$. 

The $\{ d\delta(\xi_a) \}_{a=1 \cdots dim \mathfrak{g}}$ span a distribution $\mathfrak{W}$ over $M$ as a subbundle of $\Lambda^n(T^*M)$. Let $\mathfrak{W}_m$ where $m\varepsilon M$ be a typical fibre of the distribution $\mathfrak{W}$, \hspace{1pt} then $\mathfrak{W}_m=\langle d\delta(\xi_a)(m)\rangle_{a=1 \cdots dim \mathfrak{g}}$. We can define a Lie bracket on the fibres of $\mathfrak{W}$ as follows: \begin{equation}\{d\delta(\xi_a),d\delta(\xi_b)\}^\mathfrak{W}:=d\{\delta(\xi_a),\delta(\xi_b)\}\end{equation} 
{\begin{prop} $(\mathfrak{W},\{-,-\}^\mathfrak{W})$ carries a representation of the Lie Algebra $\mathfrak{g}$. \end{prop}}
\begin{proof} Let $\mathfrak{w}_a:=d\delta(\xi_a)$. By definition $\{\mathfrak{w}_a,\mathfrak{w}_b\}^\mathfrak{W}:=d\{\delta(\xi_a),\delta(\xi_b)\}=d [C^d_{ab}\delta(\xi_c)]=C^d_{ab}d\delta(\xi_c)=:C^d_{ab}\mathfrak{w}_c$.
The Jacobi identity follows from $C^d_{[ab|}C^e_{d|f]}=0$.\end{proof}

In this subsection we will obtain the observables on the constraint multisymplectic submanifold, $\mathcal{C}$. In \cite{KS} the starting point was to identify $\Lambda^*(\mathfrak{g})\otimes C^\infty(N)$ (for a symplectic manifold $N$) together with a certain nilpotent derivation as a  Koszul complex. The analogue of the commutative algebra  $C^\infty(N)$ for first order field theories is the graded Poisson-Leibniz algebra $\Lambda^*(\hftv)$.  The contrast between the anti-commutivity of the wedge product of two elements of the Lie algebra and the commutative product of functions ensured that the required Koszul differential was nilpotent \cite{KS}. The graded-commutivity of products of elements of $\Lambda^*(\hftv)$ means that $\Lambda^*(\mathfrak{g})\wedge \Lambda^*(\hftv)$ is an  unsuitable starting point. {\it Replacing $\mathfrak{g}$ by $\mathfrak{W}$ and considering the exterior product  $\Lambda^*(\hftv)\wedge\Lambda^*(\mathfrak{W})$ we are able to form a Koszul complex giving the required resolution. }

Note that the graded Poisson-Leibniz algebra of observables $\Lambda^*(\hftv)$ satisfies the  identity \[\Lambda^*(\hftv)\cong\underset{r=0}{\overset{n-1}{\bigotimes }}\Lambda^*(\Tilde{\mathcal{H}}^p(M))\] 

Then, \[\Lambda^*(\hftv)\wedge\Lambda^*(\mathfrak{W})\cong\underset{r=0}{\overset{n-2}{\bigotimes }}\Lambda^*(\Tilde{\mathcal{H}}^r(M))\otimes\Lambda^*(\Tilde{\mathcal{H}}^{n-1}(M))\wedge \Lambda^*(\mathfrak{W}).\] Let $K^{*,*}:=\Lambda^*(\Tilde{\mathcal{H}}^{n-1}(M))\wedge  \Lambda^*(\mathfrak{W})$. Now note that if $\mathfrak{d}$ is a nilpotent derivation on $K^{*,*}$ then $\mathfrak{d}^{\text{total}}:=(\underset{r=0}{\overset{n-2}{\bigotimes }}\mathbb{I})\otimes\mathfrak{d}$ is a nilpotent derivation on $\Lambda^*(\hftv)\wedge \Lambda^*(\mathfrak{W})$.

We shall define a Koszul differential on $K^{*,*}$ and via the extension on the whole of $\Lambda^*(\hftv)\wedge\Lambda^*(\mathfrak{W})$. The special role played here by the Hamiltonian (n-1)-forms is a consequence of the special emphasis they receive in the definition of the covariant momentum mapping.
{\begin{defn} $(K^{*,*},\circ):=(\Lambda^*(\Tilde{\mathcal{H}}^{n-1}(M))\wedge \Lambda^*(\mathfrak{W}),\circ)$ is a bi-graded algebra with graded product of elements $(\overset{p}{\mathfrak{F}}\wedge\overset{r}{\mathfrak{u}})\varepsilon K^{p,r}$ and $ (\overset{q}{\mathfrak{H}}\wedge\overset{s}{\mathfrak{v}})\varepsilon K^{q,s}$ defined by
\begin{equation*}
(\overset{p}{\mathfrak{F}}\wedge\overset{r}{\mathfrak{u}})\circ(\overset{q}{\mathfrak{H}}\wedge\overset{s}{\mathfrak{v}}):=(\overset{p}{\mathfrak{F}}\wedge\overset{r}{\mathfrak{u}})\wedge(\overset{q}{\mathfrak{H}}\wedge\overset{s}{\mathfrak{v}})=(-1)^{qp(n-1)^2 +rsn^2}(\overset{q}{\mathfrak{H}}\wedge\overset{s}{\mathfrak{v}})\circ(\overset{p}{\mathfrak{F}}\wedge\overset{r}{\mathfrak{u}})
\end{equation*}
\end{defn}}
{\begin{defn} We define a differential on $\mathfrak{d}_k$ on $(K^{*,*},\circ)$ as
\begin{equation*}\mathfrak{d}_k:=(\mathbb{I}\wedge\frac{\partial}{\partial\mathfrak{w}^d})\circ(\delta(\xi_d)\wedge\mathbb{I})\end{equation*} acting on  the right of $K^{*,*}$. The action on generators 
\begin{equation*}
\begin{split}
(\mathbb{I}\wedge\mathfrak{w}_a)\varepsilon K^{1,1} &\text { and } (\overset{1}{\mathfrak{F}}\wedge\mathbb{I})\varepsilon K^{1,0}  \\
\text {is } 
(\overset{1}{\mathfrak{F}}\wedge\mathbb{I})\mathfrak{d}_k:=0 &\text { and } (\mathbb{I}\wedge\mathfrak{w}_a)\mathfrak{d}_k:=(\delta(\xi_d)\wedge\mathbb{I}).
\end{split}
\end{equation*}
\end{defn}}
{\begin{lem} Let $\partial^{\mathfrak{w}}_d:=\frac{\partial}{\partial\mathfrak{w}^d}$.
$\partial^{\mathfrak{w}}_d$ is a derivation on $\Lambda^*(\mathfrak{W})$ of degree -1.\end{lem}}
\begin{proof}
Consider $\mathfrak{w}_a\wedge\mathfrak{w}_b\varepsilon \Lambda^2 (\mathfrak{W})$, then
\begin{equation*}
\begin{split}
[\mathfrak{w}_a\wedge \mathfrak{w}_b]\partial^{\mathfrak{w}}_d=&   [d(\mathfrak{w}_a\wedge \mathfrak{w}_b)]\lhook \partial^{\mathfrak{w}}_d \\
=&  ((d\mathfrak{w}_a)\wedge\mathfrak{w}_b +(-1)^{n}\mathfrak{w}_a\wedge(d\mathfrak{w}_b)) \lhook \partial^{\mathfrak{w}}_d\\
=& \delta^a_d \mathfrak{w}_b +(-1)^{n}\mathfrak{w}_a\delta^b_d
\end{split}
\end{equation*}
It then follows that for $\mathfrak{v}\varepsilon\Lambda^s(\mathfrak{W})$ and $\mathfrak{u}\varepsilon\Lambda^r(\mathfrak{W})$ that $[\mathfrak{v}\wedge\mathfrak{u}]\partial^{\mathfrak{w}}_d=([\mathfrak{v}]\partial^{\mathfrak{w}}_d)\wedge\mathfrak{u} + (-1)^{sn}\mathfrak{v}\wedge ([\mathfrak{u}]\partial^{\mathfrak{w}}_d)$.
\end{proof}
{\begin{lem} $\mathfrak{d}_k$ is a nilpotent (of index 1) derivation of degree -1 on $K^*:=\underset{r}{\bigoplus}K^{*,r}:=\underset{r}{\bigoplus}\Lambda^*(\Tilde{\mathcal{H}}^{n-1}(M)) \wedge\Lambda^r(\mathfrak{W})$.\end{lem}}
\begin{proof}
 That $\mathfrak{d}_k$ is a derivation of degree -1 follows from the previous Lemma. We shall give an inductive proof that $\mathfrak{d}_k$ is nilpotent. Clearly $\mathfrak{d}_k^2$ vanishes on the generators. More generally let $(\overset{p}{\mathfrak{F}}\wedge\overset{r}{\mathfrak{u}})\varepsilon K^{p,r}$ and $ (\overset{q}{\mathfrak{H}}\wedge\overset{s}{\mathfrak{v}})\varepsilon K^{q,s}$. Assuming that $(\overset{p}{\mathfrak{F}}\wedge\overset{r}{\mathfrak{u}})\mathfrak{d}_k^2=0$ and $(\overset{q}{\mathfrak{H}}\wedge\overset{s}{\mathfrak{v}})\mathfrak{d}_k^2=0$
\begin{equation*}
\begin{split}
[(\overset{p}{\mathfrak{F}}\wedge\overset{r}{\mathfrak{u}})\circ(\overset{q}{\mathfrak{H}}\wedge\overset{s}{\mathfrak{v}})]\mathfrak{d}_k^2=& [(\overset{p}{\mathfrak{F}}\wedge\overset{r}{\mathfrak{u}})\circ([\overset{q}{\mathfrak{H}}\wedge\overset{s}{\mathfrak{v}}]\mathfrak{d}_k) \\
+&(-1)^{sn}(-1)^{q(n-1)^2}([\overset{p}{\mathfrak{F}\wedge}\overset{r}{\mathfrak{u}}]\mathfrak{d}_k)\circ(\overset{q}{\mathfrak{H}}\wedge\overset{s}{\mathfrak{v}})]\mathfrak{d}_k \\
=&(-1)^{(s-1)n}(-1)^{(q-1)(n-1)^2}([\overset{p}{\mathfrak{F}}\wedge\overset{r}{\mathfrak{u}}]\mathfrak{d}_k)\circ([\overset{q}{\mathfrak{H}}\wedge\overset{s}{\mathfrak{v}}]\mathfrak{d}_k) \\
+&(-1)^{sn}(-1)^{q(n-1)^2}([\overset{p}{\mathfrak{F}\wedge}\overset{r}{\mathfrak{u}}]\mathfrak{d}_k)\circ([\overset{q}{\mathfrak{H}}\wedge\overset{s}{\mathfrak{v}}]\mathfrak{d}_k)
\end{split}
\end{equation*}
Note that $(s-1)n+(q-1)(n-1)^2-(sn+q(n-1)^2)=(n-1)(n-2)-1$ is an odd number. Therefore one of the powers must be even and the other must be odd. We therefore have $[(\overset{p}{\mathfrak{F}}\wedge\overset{r}{\mathfrak{u}})\circ(\overset{q}{\mathfrak{H}}\wedge\overset{s}{\mathfrak{v}})]\mathfrak{d}_k^2=0$. The statement then follows  by induction.
\end{proof}
The complex $(\Lambda^*(\hftv)\wedge\Lambda^*(\mathfrak{W}),\mathfrak{d}_K)$ where $\mathfrak{d}_K:=(\underset{r=0}{\overset{n-2}{\bigotimes }}\mathbb{I})\otimes\mathfrak{d}_k$ forms the following long exact sequence:
\begin{equation*}
\begin{split}
 0\longrightarrow & \Lambda^*(\Tilde{\mathcal{H}}^*(M))\wedge \Lambda^{dim\mathfrak{G}}(\mathfrak{W})
\overset{\mathfrak{d}_K^{{dim\mathfrak{G}}}}{\longrightarrow} \Lambda^*(\Tilde{\mathcal{H}}^*(M))\wedge \Lambda^{dim\mathfrak{G}-1}(\mathfrak{W})
\overset{\mathfrak{d}_K^{{dim\mathfrak{G}}-1}}{\longrightarrow}\dots \\
& \dots \overset{\mathfrak{d}_K^2}{\longrightarrow} \Lambda^*(\Tilde{\mathcal{H}}^*(M))\wedge \Lambda^1(\mathfrak{W})
\overset{\mathfrak{d}_K^1}{\longrightarrow} \Lambda^*(\Tilde{\mathcal{H}}^*(M)) \wedge\Lambda^0(\mathfrak{W})
\overset{\mathfrak{d}_K^0}{\longrightarrow} 0
\end{split}
\end{equation*}
By virtue of this long exact sequence and the action of the differential on the generators we identify the complex $(\Lambda^*(\hftv)\wedge\Lambda^*(\mathfrak{W}),\mathfrak{d}_K)$ as a Koszul complex. Note that 
\begin{gather*}
\text{ker}{\mathfrak{d}_K^0}=\Lambda^*(\Tilde{\mathcal{H}}^*(M)) \qquad
\text{Im}{\mathfrak{d}_K^1}=\Lambda^*(\Tilde{\mathcal{H}}^*(M))\wedge \delta[\mathfrak{g}] \\
\Longrightarrow \quad H^0_{\mathfrak{d}_K}[(\Lambda^*(\hftv)\wedge\Lambda^*(\mathfrak{W})  ]:=\frac{\text{ker}{\mathfrak{d}_K^0}}{\text{Im}{\mathfrak{d}_K^1}}=\frac{\Lambda^*(\hftv)}
{\Lambda^*(\hftv)\wedge \delta[\mathfrak{g}]}
\end{gather*}
From a classical result in homological algebra \cite{Lang} the higher homology groups vanish. That is
{\begin{thm} The homology of the complex $(\Lambda^*(\hftv)\wedge\Lambda^*(\mathfrak{W}),\mathfrak{d}_K)$  is 
\begin{equation}
H^p_{\mathfrak{d}_K}[ (\Lambda^*(\hftv)\wedge\Lambda^*(\mathfrak{W})  ]=
\begin{cases}
0 & \text{if $p>0$} \\
\frac{\Lambda^*(\hftv)}
{\Lambda^*(\hftv)\wedge \delta[\mathfrak{g}]} & \text{if $p=0$}
\end{cases}
\end{equation}
\end{thm}
We have therefore obtained the algebra of observables on the constraint multisymplectic submanifold as a Koszul resolution.
\subsection{The Chevalley-Eilenberg Complex}
The first part of the homological treatment of Marsden-Weinstein multisymplectic  reduction was carried out in the subsection above where we obtained the observables on the constraint submanifold of $(M,\Omega)$ defined to be $\mathbb{J}^{-1}(0)$. In this section we shall complete the reduction process by obtaining the observables on the reduced multisymplectic manifold, $\mathcal{B}$. They are the $\mathfrak{g}$-invariant elements of $H_{\mathfrak{d}_K}^0(\Lambda^*(\hftv)\wedge\Lambda^*(\mathfrak{W}))$. We obtain them by consideration of the cohomology of $\mathfrak{g}$ with values in the Koszul complex and its homology groups. This requires that the Koszul  homology groups be $\mathfrak{g}$-modules. We therefore begin by proving that the Koszul complex is a right $\mathfrak{g}$-module and then by virtue of the fact that the Chevalley-Eilenberg differential and $\mathfrak{d}_K$ commute, the module structure descends to the homology. We begin by proving an identity which we shall use to construct a representation of $\mathfrak{g}$ as an endomorphism of $\Lambda^*(\hftv)\wedge\Lambda^*(\mathfrak{W})$. 
{\begin{lem}
Let $G,H\varepsilon\hft$, $\mathfrak{E}\varepsilon\Lambda^*(\hftv)$ then
\begin{equation}
\{\mathfrak{E},\{G,H\}\}=\{\{\mathfrak{E},G\},H\}-\{\{\mathfrak{E},H\},G\}
\end{equation}\end{lem}}
\begin{proof}
The identity is true for all $\mathfrak{E}\varepsilon\Lambda^1(\hftv)$  is an immediate consequence on the graded Jacobi identity for $(\hftv,\{\cdot,\cdot\})$. We shall give a proof by induction. Let us assume that the identity is true for 
$\mathfrak{F}\varepsilon\Lambda^p(\hftv)$ and show that it remains true for$\mathfrak{E}:=\mathfrak{F}\wedge\mathfrak{K}$, where $\mathfrak{E}\varepsilon\Lambda^q(\hftv)$. By virtue of the right Leibniz rule, and computing each term one finds:
\begin{equation*}
\begin{split}
\{\mathfrak{E},\{G,H\}\}= & \{\mathfrak{F}\wedge\mathfrak{K},\{G,H\}\}     \\
= & \mathfrak{F}\wedge \{\mathfrak{K},\{G,H\} \}+ (-1)^{\mathfrak{K}(g_G + g_H)}\{\mathfrak{F}\{G,H\}\}\wedge\mathfrak{K} \\
\{\{\mathfrak{E},G\},H\}=&\{\{\mathfrak{F}\wedge\mathfrak{K},G\},H\} \\
=& \{\mathfrak{F}\wedge\{\mathfrak{K},G\},H\} +(-1)^{\mathfrak{K}(g_G)}\{\{\mathfrak{F},G\}\wedge\mathfrak{K},H\} \\
=& \mathfrak{F}\wedge\{\{\mathfrak{K},G\},H\} +(-1)^{(\mathfrak{K} + G -n+1)g_H}\{\mathfrak{F},H\}\wedge\{\mathfrak{K},G\}\\
+&(-1)^{\mathfrak{K}(g_G)}\{\mathfrak{F},G\}\wedge\{\mathfrak{K},H\} +
(-1)^{\mathfrak{K}(g_G)}(-1)^{\mathfrak{K}(g_H)}\{\{\mathfrak{F},H\},G\}\wedge\mathfrak{K} \\
\{\{\mathfrak{E},H\},G\}=&\{\{\mathfrak{F}\wedge\mathfrak{K},H\},G\} \\
=&\{\mathfrak{F}\wedge\{\mathfrak{K},H\},G\} +(-1)^{\mathfrak{K}g_H}\{\{\mathfrak{F}H\}\wedge\mathfrak{K},G\} \\
=&\mathfrak{F}\wedge\{\{\mathfrak{K},H\},G\} +(-1)^{(\mathfrak{K} + H -n+1)g_G}\{\mathfrak{F},G\}\wedge\{\mathfrak{K},H\} \\
+& (-1)^{\mathfrak{K}(g_H)}\{\mathfrak{F},H\}\wedge\{\mathfrak{K},G\}
+ (-1)^{\mathfrak{K}(g_G)}(-1)^{\mathfrak{K}(g_H)}\{\{\mathfrak{F},H\},G\}\wedge\mathfrak{K}
\end{split}
\end{equation*}
The cross terms cancel since $g_H=g_G=0$ (n.b. they would not cancel for forms of any degree so that this result is special to (n-1)-forms). Furthermore the remaining terms become 
\begin{equation*}
\begin{split}
&\{\mathfrak{E},\{G,H\}\}-\{\{\mathfrak{E},G\},H\}+\{\{\mathfrak{E},H\},G\}\\
&=\mathfrak{F}\wedge [\{\mathfrak{K},\{G,H\} \}-\{\{\mathfrak{K},G\},H\}+\{\{\mathfrak{K},H\},G\}] \\
&(-1)^{\mathfrak{K}(g_G + g_H)}[\{\mathfrak{F}\{G,H\}\}-\{\{\mathfrak{F},G\},H\}+\{\{\mathfrak{F},H\},G\}]\wedge\mathfrak{K}
\end{split}
\end{equation*}
So that the statement is true by  induction.
\end{proof}
We wish to show  that $\Lambda^*(\hftv)\wedge\Lambda^*(\mathfrak{W})$ is a right $\mathfrak{g}$-module. We therefore need to show that $\exists$ a Lie algebra morphism: 
\begin{equation*}
\varrho:\mathfrak{g}\rightarrow End(\Lambda^*(\hftv)\wedge\Lambda^*(\mathfrak{W})) \isto \varrho:\xi \longmapsto\varrho(\xi)
\end{equation*}
for all $\xi\varepsilon\mathfrak{g}$ such that the following properties hold:\newline
1.$\Lambda^*(\hftv)\wedge\Lambda^*(\mathfrak{W})     \times \mathfrak{g}\rightarrow \Lambda^*(\hftv)\wedge\Lambda^*(\mathfrak{W}) $\newline $\isto (\mathfrak{F}\wedge \mathfrak{w},\xi)\longmapsto [\mathfrak{F}\wedge \mathfrak{w}]\varrho(\xi) \text{ for all }\mathfrak{F}\wedge \mathfrak{w} \varepsilon\Lambda^*(\hftv\wedge)\Lambda^*(\mathfrak{W})$\newline
2. $[\mathfrak{F}\wedge \mathfrak{w}]\varrho(\xi)$ is linear in $\xi$ and $\mathfrak{F}$.\newline
3. $\varrho([\xi,\zeta])=[\varrho(\xi),\varrho(\zeta)]$\newline
{\begin{prop} $\Lambda^*(\hftv)\wedge \Lambda^*(\mathfrak{W})$ is a right  $\mathfrak{g}$-module.\end{prop}}
\begin{proof}
Let us consider each point in tern:\newline
1. We define a representation $\varrho$ on $\Lambda^*(\hftv)\otimes\Lambda^*(\mathfrak{W})$ by:
\begin{gather*}
[\mathfrak{F}\wedge  \mathbb{I}]\varrho (\xi_a):=\{\mathfrak{F},\delta(\xi_a)\}\wedge  \mathbb{I}
\quad [\mathfrak{F}\wedge\mathfrak{w}_b ]\varrho (\xi_a) :=\mathfrak{F}\wedge \{\mathfrak{w}_b,\mathfrak{w}_a\}^\mathfrak{W}+  \{\mathfrak{F},\delta(\xi_a)\}\wedge\mathfrak{w}_b
\end{gather*}
2. Linearity in  the $\xi$ follows from the linearity of the Lie bracket $\{\cdot,\cdot\}^\mathfrak{W}$. Linearity in the $\mathfrak{F}$ follows from linearity of the  Leibniz bracket $\{\cdot,\cdot\}$.\newline
3. Now, $[\mathfrak{w}_c\wedge \mathfrak{F}]\varrho([\xi_a,\xi_b]):=\mathfrak{F}\wedge \{\mathfrak{w}_c,\{\mathfrak{w}_a,\mathfrak{w}_b\}\}^\mathfrak{W}+  \{\mathfrak{F},\{\delta(\xi_a),\delta(\xi_b)\}\}\wedge\mathfrak{w}_c$.\newline
  Since $\delta(\xi)\varepsilon \hft$, one has  $\delta([\xi,\zeta])=\{\delta(\xi),\delta(\zeta)\}$.\newline
Then by virtue of the last Lemma and the  Jacobi identity for $(\mathfrak{W},\{\cdot,\cdot\}^\mathfrak{W})$  we have \begin{equation*}
\begin{split}
[\mathfrak{F}\wedge\mathfrak{w}_c ]\varrho([\xi_a,\xi_b])=&\{\{\mathfrak{F},\delta(\xi_a)\},\delta(\xi_b)\}\wedge\mathfrak{w}_c -\{\{\mathfrak{F},\delta(\xi_b)\},\delta(\xi_a)\}\wedge\mathfrak{w}_c \\+&\mathfrak{F}\wedge\{\{\mathfrak{w}_c,\mathfrak{w}_a\},\mathfrak{w}_b\}^\mathfrak{W}-\mathfrak{F\wedge}\{\{\mathfrak{w}_c,\mathfrak{w}_b\},\mathfrak{w}_a\}^\mathfrak{W} \\
\Longrightarrow   \qquad   \qquad           =&\{\{\mathfrak{F},\delta(\xi_a)\},\delta(\xi_b)\}\wedge\mathfrak{w}_c -\{\{\mathfrak{F},\delta(\xi_b)\},\delta(\xi_a)\}\wedge\mathfrak{w}_c \\
+& \{\mathfrak{F}, \delta(\xi_a)\}\wedge\{\mathfrak{w}_c,\mathfrak{w}_b\}^\mathfrak{W}+ \{\mathfrak{F}, \delta(\xi_b)\}\wedge\{\mathfrak{w}_c,\mathfrak{w}_a\}^\mathfrak{W}\\
+& \mathfrak{F}\wedge\{\{\mathfrak{w}_c,\mathfrak{w}_a\},\mathfrak{w}_b\}^\mathfrak{W}-\mathfrak{F}\wedge\{\{\mathfrak{w}_c,\mathfrak{w}_b\},\mathfrak{w}_a\}^\mathfrak{W}    \\
-& \{\mathfrak{F}, \delta(\xi_a)\}\wedge\{\mathfrak{w}_c,\mathfrak{w}_b\}^\mathfrak{W}- \{\mathfrak{F}, \delta(\xi_b)\}\wedge\{\mathfrak{w}_c,\mathfrak{w}_a\}^\mathfrak{W} \\
\Longrightarrow   \qquad       \qquad          =&[\mathfrak{F}\wedge\mathfrak{w}_c ][\varrho(\xi_a),\varrho (\xi_b)]
\end{split}
\end{equation*}
as required.
\end{proof}
Note that we have defined the representation to act to the right precisely so that we could use Lemma 3.7. to prove the homomorphism property. The parts of the two (n-1)-forms is  played doubly by the covariant Noether current. The fact that the covariant Noether currents are defind to be (n-1)-forms thus plays an essential role here.  As a consequence of the non-commutivity of the Leibniz bracket if we had defined the representation to act to the left then neither would there be as simple an identity as in Lemma 3.7. nor would the homomorphism property have been true.

Now we introduce the Chevalley-Eilenberg complex constructed by tensoring to the right of  the Koszul complex by the exterior algebra over the dual of the Lie algebra. Let $\{\alpha^a\}^{a=1\cdots dim\mathfrak{G}}$ be a basis of  $\mathfrak{g}^*$. Geometrically we may identify the $\{\alpha^a\}^{a=1\cdots dim\mathfrak{G}}$ with the coframe basis of the cotangent bundle to the isotropic submanifolds foliating $\mathcal{C}$. The coframe is therefore defined by $\alpha^aX(\xi_b)=\delta^a_b$. The basis  satisfies the {\it Maurer-Cartan structural equations}, namely $d\alpha^a=\frac{1}{2}C^a_{bc}\alpha^b \wedge \alpha^c$. 
{\begin{defn} let $(L,\circledcirc):=(\underset{r,s}{\bigoplus}L^{r,s}:=\underset{r,s}{\bigoplus}\Lambda^*(\hftv)\wedge\Lambda^r(\mathfrak{W}) \otimes\Lambda^s(\mathfrak{g}^*),\circledcirc)$, where the binary operation  $\circledcirc$ is defined to act as follows 
\begin{equation*}(\mathfrak{F}\wedge\mathfrak{w}\otimes\mu)\circledcirc(\mathfrak{E}\wedge\Acute{\mathfrak{w}}\otimes\Acute{\mu}):=(\mathfrak{F}\wedge\mathfrak{E}\wedge\mathfrak{w}\wedge\Acute{\mathfrak{w}}\otimes\mu\wedge\Acute{\mu})\end{equation*}
for all $\mathfrak{F},\mathfrak{G}\varepsilon\Lambda^*(\hftv), \quad
\mathfrak{w},\Acute{\mathfrak{w}}\varepsilon\Lambda^*(\mathfrak{W}) \text{ and }
\mu,\Acute{\mu}\varepsilon\Lambda^*(\mathfrak{g}^*)$. Let $d$ be a derivation of degree 1 defined by 
\begin{equation}
\begin{split}
&d: \Lambda^*(\hftv)\wedge\Lambda^p(\mathfrak{W})\otimes \Lambda^q(\mathfrak{g}^*)\rightarrow \Lambda^*(\hftv)\wedge\Lambda^p(\mathfrak{W})\otimes \Lambda^{q+1}(\mathfrak{g}^*) \hspace{5pt}\\
&::\hspace{5pt} 
(\mathfrak{F}\wedge\mathfrak{w}\otimes\mu)d:=\mathfrak{F}\wedge\mathfrak{w}\otimes(d[\mu]) + (-1)^{deg \mu}\underset{a=1\cdots n}{\Sigma}[\mathfrak{F}\wedge\mathfrak{w}]\varrho[\xi_a]\otimes\alpha^a\wedge\mu
\end{split}
\end{equation}
\end{defn}}
{\begin{lem} The derivation $d$ is nilpotent (of index 1).\end{lem}}
\begin{proof}
By direct computation
\begin{equation}
\begin{split}
(\mathfrak{F}\wedge\mathfrak{w}\otimes\mu)d^2=&\mathfrak{F}\wedge\mathfrak{w}\otimes(d^2[\mu]) + (-1)^{deg \mu+1}\underset{a=1\cdots n}{\Sigma}[\mathfrak{F}\wedge\mathfrak{w}]\varrho[\xi_a]\otimes\alpha^a\wedge (d[\mu]) \\
&(-1)^{deg \mu}\underset{a=1\cdots n}{\Sigma}[\mathfrak{F}\wedge\mathfrak{w}]\varrho[\xi_a]\otimes\alpha^a\wedge (d[\mu])\\
&(-1)^{2deg \mu+1}\underset{a=1\cdots n,b=1\cdots n}{\Sigma}[[\mathfrak{F}\wedge\mathfrak{w}]\varrho[\xi_a]]\varrho[\xi_b]\otimes\alpha^b\wedge\alpha^a\wedge \mu\\
&(-1)^{2deg \mu}\underset{a=1\cdots n}{\Sigma}[\mathfrak{F}\wedge\mathfrak{w}]\varrho[\xi_a]\otimes(d[\alpha^a])\wedge \mu
\end{split}
\end{equation}
The second two terms cancel and the first term vanishes immediately. 
Now 
\begin{equation}
\alpha^a \wedge \alpha^b \otimes \varrho(\xi_b) \varrho(\xi_a)=
\frac{1}{2}\alpha^a \wedge \alpha^b[\varrho(\xi_b),\varrho(\xi_a)]
=\frac{1}{2}\alpha^a \wedge \alpha^b C^c_{ba}\varrho(\xi_c)
\end{equation}
by virtue of which and the Maurer-Cartan structural equations the remaining two terms cancel.
\end{proof}
{\begin{lem} The derivations $d$ and $\mathfrak{d}_K$ commute.\end{lem}}

\begin{proof}
This follows by virtue of the fact that $\varrho$ and $\mathfrak{d}_K$ commute.
Now, using the right graded Leibniz identity for $\Lambda^*(\hftv)$ in the third line one has:
\begin{equation}
\begin{split}
[(\mathfrak{F}\wedge\mathfrak{w}_a)]\mathfrak{d}_K \varrho(\xi_b):=&
\mathfrak{F}\wedge\delta(\xi_a)\wedge \mathbb{I} \\
:=& \{\mathfrak{F}\wedge\delta(\xi_a),\delta(\xi_b)\} \\
=&\mathfrak{F}\wedge\{\delta(\xi_a),\delta(\xi_b)\} +
\{\mathfrak{F},\delta(\xi_b)\}\wedge\delta(\xi_a) \\
=&[\mathfrak{F}\wedge\{\mathfrak{w}_a,\mathfrak{w}_b\}^\mathfrak{W} +
\{\mathfrak{F},\delta(\xi_b)\}\wedge\mathfrak{w}_a]\mathfrak{d}_K\\
=:&[(\mathfrak{F}\wedge\mathfrak{w}_a)]\varrho(\xi_b)\mathfrak{d}_K
\end{split}
\end{equation}
\end{proof}
If we had chosen the representation to act to the left instead of the right, $d$ and $\mathfrak{d}_K$ would have failed to commute. Again we see that in  extending  the structures from  classical  mechanics and the  Poisson algebra of observables to the field theoretic setting of the graded Poisson-Leibniz algebra we must make a judicious use of the left Leibniz rule and the graded Loday identity.

We know that $H^p_{\mathfrak{d}_K}[ (\Lambda^*(\hftv)\wedge\Lambda^*(\mathfrak{W})  ]$ is a right \lalg -module, as the above lemma tells us that the module structure of $\Lambda^*(\hftv)\wedge\Lambda^*(\mathfrak{W})$ descends to the level of homology. In particular $H^0_{\mathfrak{d}}[\Lambda^*(\hftv)\wedge\Lambda^*(\mathfrak{W})  ]$ is a right \lalg -module. We therefore have the following result.
{\begin{thm} The zeroth level Lie algebra cohomology with values in the zeroth Koszul homology is the algebra of observables of the reduced multisymplectic manifold $\mathcal{B}$,
\begin{equation}
\begin{split}
H^0_d(\mathfrak{g},&H^0_{\mathfrak{d}_K}[ (\Lambda^*(\hftv)\wedge\Lambda^*(\mathfrak{W})] )\\
&=\mathfrak{g}\text{  invariants of } 
H^0_{\mathfrak{d}_K}[ (\Lambda^*(\hftv)\wedge\Lambda^*(\mathfrak{W})] \\
&= \mathfrak{g}\text{  invariants of }  \frac{\Lambda^*(\hftv)}
{\Lambda^*(\hftv)\wedge \delta[\mathfrak{g}]  }
\end{split}
\end{equation}
\end{thm}}
\begin{proof}
Ker$[d^0]$=the inverse image of the trivial endomorphisms of the right \lalg -module $H^0_{\mathfrak{d}_K}[ (\Lambda^*(\hftv)\wedge\Lambda^*(\mathfrak{W})]$ so that
 \begin{equation} Ker[d^0]= \text{the \lalg-invariants of } H^0_{\mathfrak{d}_K}[ (\Lambda^*(\hftv)\wedge\Lambda^*(\mathfrak{W})]\end{equation}
furthermore, Im$[i]$= 0. The zeroth cohomology is then given by 
\begin{equation*}
\begin{split}
\Longrightarrow H^0_d(\mathfrak{g},&H^0_{\mathfrak{d}_K}[ (\Lambda^*(\hftv)\wedge\Lambda^*(\mathfrak{W})] )= \frac{\text{Ker}[d^0]}{\text{Im}[i]}\\=&\text{the \lalg-invariants of }\frac{\Lambda^*(\hftv)}
{\Lambda^*(\hftv)\wedge \delta[\mathfrak{g}]  }
\end{split}
\end{equation*}
\end{proof}
The commutivity of the Koszul differential and the Chevalley-Eilenberg differential means that they form a double complex. Because the Koszul homology vanishes for at all levels higher than the zeroth this means that the spectral sequence degenerates (for more details see \cite{kimura}), and we have the following result: The observables on $\mathcal{B}$ may be obtained by considering the homology of the total differential \begin{equation}\mathcal{D}=d +(-1)^p(1\otimes \mathfrak{d}_K)\end{equation} on the bi-complex $\Lambda^*(\hftv)\wedge\Lambda^*(\mathfrak{W})\otimes \Lambda^*(\mathfrak{g}^*)$, whence
\begin{equation*}
H^0_{\mathcal{D}}(\Lambda^*(\hftv)\wedge\Lambda^*(\mathfrak{W})\otimes \Lambda^*(\mathfrak{g}^*))= H^0_d(\mathfrak{g}, H_{\mathfrak{d}}^0(\Lambda^*(\hftv)\wedge\Lambda^*(\mathfrak{W})))
\end{equation*}
The differential $D$ is the classical BRST differential.

\section{Conclusion}

We have advocated the graded Poisson-Leibniz algebra of observables, in the multisymplectic framework, as the appropriate generalisation of the  Poisson(-Lie) algebra of observables, in the symplectic framework \cite{kan1,kan2,kan3}. When n=0 we regain the symplectic context. We have found that   in  extending  the structures from  the symplectic setting  of the  Poisson algebra of observables to the multisymplectic setting of the graded Poisson-Leibniz algebra we must make a judicious use of definitions so that we may utilise the left Leibniz rule and the graded Loday identity. With this observation, and by virtue of the introduction of a certain distribution of n-forms carrying a representation of the Lie algebra in its fibres, we have thus been able to generalise the homological description of symplectic reduction to the multisymplectic case.

Various papers in the multisymplectic literature have assumed that the Lie algebra of Hamiltonian (n-1)-forms is the appropriate candidate for the algebra of observables. However as this paper has shown in order give a homological description of reduction, a necessary prerequisite to the classical BRST symmetry, in the multisymplectic framework  we must introduce a product on forms. The ensuing algebraic structure is not a Lie algebra but shares enough structure in common with a Poisson algebra to be of use.

In an accompanying paper \cite{jdg2} we  develop a covariant Hamiltonian formulation of the classical BRST symmetry. The homological description of Marsden-Weinstein multisymplectic reduction expounded within this paper is the prerequisite foundational step towards a multisymplectic formulation of the classical BRST symmetry. We develop there the multisymplectic analogs of the Lagrange-d'Alembert Hamiltonian formalism and introduce certain graded multisymplectic manifolds which form the geometric foundations of a multisymplectic treatment of the BRST symmetry. We shall show that the action of the total differential $D$  on the complex $\Lambda^*(\hftv)\wedge\Lambda^*(\mathfrak{W})\otimes \Lambda^*(\mathfrak{g}^*)$ is equivalent to the Leibniz derivation of a  Grassmann-odd generalised Hamiltonian (n-1)-form on a subalgebra of the bi-graded Poisson-Leibniz algebra of observables on a  graded-multisymplectic manifold. We shall also discover that in the geometric formalism the introduction of the Poisson-Leibniz algebra of observables is indispensable. 
\vspace{10pt}

{\scshape {Acknowledgements}} The author would like to thank Alice Rogers for her encouragement with this article.

\end{document}